%% file: main.tex
\documentclass[twocolumn]{aastex631}
\usepackage{soul}

\newcommand{\webb}{JWST}
\newcommand{\jwst}{JWST}
\newcommand{\tess}{TESS}
\newcommand{\mum}{\ifmmode{\rm \mu m}\else{$\mu$m}\fi}
\newcommand{\lightkurve}{{{\fontfamily{lmtt}\selectfont lightkurve}}}
\newcommand{\tesslocalize}{{{\fontfamily{lmtt}\selectfont
\newcommand{\Msun}      {\>{\rm M_{\odot}}}
TESS\_Localize}}}

\shorttitle{TESS Variability of JWST Calibration Stars}
\shortauthors{Mullally et al.}

\graphicspath{{./}{figures/}}

\begin{document}

\title{Searching for TESS Photometric Variability of Possible JWST Spectrophotometric Standard Stars}

\author[0000-0001-7106-4683]{Susan E.\ Mullally}
\affiliation{Space Telescope Science Institute, 3700 San Martin Dr.,
  Baltimore, MD 21218, USA}
\email{smullally@stsci.edu}

\author[0000-0003-4520-1044]{G.~C.~Sloan}
\affiliation{Space Telescope Science Institute, 3700 San Martin Dr., Baltimore, MD 21218, USA}
\affiliation{Department of Physics and Astronomy, University of North
  Carolina, Chapel Hill, NC 27599-3255, USA}

\author[0000-0001-5941-2286]{J.~J.~Hermes}
\affiliation{Department of Astronomy \& Institute for Astrophysical Research, Boston University, 725 Commonwealth Ave., Boston, MA 02215, USA}

\author[0000-0001-6322-2784]{Michael Kunz}
\affiliation{Space Telescope Science Institute, 3700 San Martin Dr.,
  Baltimore, MD 21218, USA}
\affiliation{Albert Einstein High School, Kensington, MD, 20895 }

\author[0000-0001-5473-856X]{Kelly Hambleton}
\affiliation{Villanova University, 800 Lancaster Avenue, Villanova, PA 19850, USA}

\author[0000-0001-9806-0551]{Ralph Bohlin}
\affiliation{Space Telescope Science Institute, 3700 San Martin Dr., Baltimore, MD 21218, USA}

\author[0000-0003-0556-027X]{Scott W. Fleming}
\affiliation{Space Telescope Science Institute, 3700 San Martin Dr.,
  Baltimore, MD 21218, USA}
  
\author[0000-0001-5340-6774]{Karl D.\ Gordon}
\affiliation{Space Telescope Science Institute, 3700 San Martin Dr., Baltimore, MD 21218, USA}

\author[0000-0002-3491-9010]{Catherine Kaleida}
\affiliation{Space Telescope Science Institute, 3700 San Martin Dr., Baltimore, MD 21218, USA}
  
\author[0000-0001-9502-3448]{Khalid Mohamed}
\affiliation{Space Telescope Science Institute, 3700 San Martin Dr.,
  Baltimore, MD 21218, USA}
\affiliation{Department of Physics \& Astronomy, Amherst College, C025 Science Center 25 East Drive, Amherst, MA 01002, USA}

\begin{abstract}
We use data from the Transiting Exoplanet Survey Satellite (\tess) to search for, and set limits on, optical to near-infrared photometric variability of the well-vetted, candidate James Webb Space Telescope (JWST) spectrophotometric standards. Our search of 37 of these candidate standards has revealed measurable periodic variability in 15 stars. The majority of those show variability that is less than half a percent; however, four stars are observed to vary photometrically, from minimum to maximum flux, by more than 1\% (the G dwarf \object{HD~38949} and three fainter A dwarfs).  Variability of this size would likely impact the error budget in the spectrophotometric calibration of the science instruments aboard \jwst.  For the 22 candidate standards with no detected variability, we report upper limits on the observed changes in flux.  Despite some systematic noise, all stars brighter than $12^{th}$\,magnitude in the TESS band show a 3$\sigma$ upper limit on the total change in brightness of less than half a percent on time scales between an hour and multiple weeks, empirically establishing their suitability as spectrophotometric standards. We further discuss the value and limits of high-cadence, high-precision photometric monitoring with TESS as a tool to vet the suitability of stars to act as spectrophotometric standards.

\end{abstract}


\section{Introduction} 
\label{sec:intro} 

The James Webb Space Telescope (\webb), an infrared space telescope with a diameter of 6.5\,m, launched at the end of 2021, promises to revolutionize many areas of astrophysics, from exoplanets to the most distant galaxies \citep{Kalirai2018,gardner2006jwst}.  In order to accomplish those goals, \webb\ will observe a sample of spectrophotometric standard stars to calibrate observations across the near- and mid-infrared (0.6--28.8~\mum).  The objective is an absolute accuracy in the observed flux of the standard stars to better than 2\% \citep[see JWST Data Absolute Flux Calibration in][]{jdox}.  A successful standard-star calibration program will not only enable the absolute calibration and the \webb\ cross-instrument calibration, but will also tie \webb\ observations to other space telescopes, such as Hubble, Spitzer, and WISE, and other ground-based observatories.

The selection of standard stars must take into account many factors that can reduce both the accuracy and the precision of the spectrophotometric calibration.  The list of possible reasons to reject potential standards during the vetting process is long, but among them, variability is always a red flag.  Variable stars should be avoided because their variations will increase the noise in the calibration data.  Even low-amplitude variability below any level of direct concern could point to more subtle concerns, such as binarity, pulsation, or strong magnetic-field activity \citep{Bohlin2014PASP126}.  Each of these issues is a reason in itself to reject a candidate, but may have otherwise gone unnoticed.  The identification of any known variable stars prior to their observation by \webb\ will give the calibration team the opportunity to investigate them more closely and potentially save valuable observing time.  Optimizing the calibration of \webb\ will contribute to the success of NASA's flagship mission.

High-cadence photometric surveys of stars, as done by missions like CoRoT \citep{corot}, Kepler \citep{Koch2010} and \tess\ \citep{Ricker2015}, have revealed that they can change brightness due to a variety of factors, and these variations can unexpectedly change in timescale or amplitude. Many of these variations are internal to the star, e.g.\ stellar pulsations \citep[e.g.][]{Berger1979deltaScuti} or spots rotating in and out of view  \citep[e.g.][]{Mcquillan2014, Balona2021}, while other variations are external, e.g.\ eclipses and transits \citep[e.g.][]{prsa2011, Thompson2018}. Time-series photometric observations have revealed many examples of atypical brightness variations where stars change brightness suddenly and in unexpected ways.  For example, Boyajian's star, KIC~8462852, was discovered using Kepler data and shows unexplained drops in the flux of the star as large as 20\% at seemingly random times \citep{Boyajian2016MNRAS}.  The nearby red supergiant Betelgeuse dimmed by more than one visual magnitude in 2019, the deepest decline reported in 50-plus years of observations \citep{Betelgeuse2021Natur594,Cotton2020RNAAS4}.  Some white dwarf stars, the classic choice for photometric standards in the UV and optical, have shown both unusual intrinsic variability from pulsations \citep[e.g.][]{Provencal2009ApJGD358,Kilic2015ApJ,Hermes2017MNRAS} and external variability caused by accretion disks \citep{Scaringi2021Nat} or disintegrating planetesimals \citep[as large as 40\%;][]{Vanderburg2015Natur,Guidry2021}.  While these examples are rare, they demonstrate that stars can vary for a variety of reasons, many of which are not predictable based on current knowledge.  

\tess\ was launched in 2018 with the purpose of identifying transiting exoplanets around nearby stars.  It photometrically observes a swath of sky covering 24 degrees by 90 degrees for a month at a time and has a bandpass of 0.6--1~\mum\ \citep{Ricker2015} that overlaps with the shortest wavelengths of the \webb\ bandpass. \tess\ is well positioned to monitor the candidate standards for \webb\ for changes in flux on time scales between minutes and weeks at a precision as fine as a few hundred parts per million \citep{Ricker2015}, an improvement in precision, cadence and coverage over previous observations to identify short-term variability in standard stars from the ground \citep[e.g.,][]{Marinoni2016}. 

As a result, \tess\ has serendipitously observed most of the candidate spectrophotometric standards for \webb\ over the course of its mission, and it will continue to do so in its extended mission.  We have examined the \tess\ data for these candidates and looked for evidence of photometric variability, and we report upper limits on any variability if none was detected.  For the 15 candidate spectrophotometric standards found to show statistically significant periodic variability, we show a light curve and periodogram and provide a few basic statistics describing the amplitude of the variations. We also discuss navigating potential pitfalls and the inherent limits of using TESS to search for various types of known variability.

\section{Candidate JWST Spectrophotometric Standard Stars} 
\label{sec:targets}

Spectrophotometric calibration requires accurate models of the stars chosen as standards, because any errors in the assumed true spectrum of a standard will propagate into the entire calibrated database \citep[e.g.][]{Cohen1992, Price2002, Sloan2015}.  Calibrating with a sample of standards will reduce the impact of problems with the model of any one star, and combining different types of standards will reduce the problems even further.  By comparing the calibration as determined from different stars, and different classes of stars, outliers can be identified and their models corrected, or, if that proves impossible, the stars can be rejected.


The \webb\ calibration will be based on three classes of standard stars:  hot stars, A dwarfs, and solar analogs (i.e., G dwarfs) \citep{Gordon2022inprep}.  It is expected that a minimum of five stars of each class will be observed during the \jwst\ mission to accomplish this task.  Comparisons of the stars will identify potential issues with models of the stars and possible outliers.  The sample examined in this paper (see Table~\ref{tab:targets}) is based on lists of candidates available in the JWST User Documentation accessed in 2020~December\footnote{\url{https://jwst-docs.stsci.edu/jwst-data-calibration-considerations/jwst-data-absolute-flux-calibration}} \citep{jdox}.

It must be stressed that the candidate list is evolving as stars are vetted and new stars are added to better cover parameter space.  \cite{Gordon2022inprep} provide details on how stars were selected and vetted.  In brief, most of the stars have been observed, usually repeatedly, in previous space missions, most notably Hubble and Spitzer.  For example, the list includes A dwarf stars reported by \citet{Reach2005} to be primary standards for IRAC on Spitzer (see those with shortened names starting with J in Table~\ref{tab:targets}). Stars identified as variables in previous missions have already been rejected.  This filter would have removed stars varying with amplitudes greater than roughly 1--2\% (e.g., \citealt{Sloan2015}).  A review of the literature for remaining candidates identifies stars with known companions and debris disks.  Spectral classifications reveal stars with unusual spectral properties that make them difficult to model and thus poor standards.  Despite the effort made, a closer look can always reveal problems that have escaped notice, and the photometric precision of TESS makes it a valuable tool in the vetting process.

\input{targetTable}

\section{TESS Observations} 
\label{sec:obs}

To study the variability of the candidate spectrophotometric standards, we use observations from \tess\ spanning Sectors 1 to 40 (years 2017--2021). When possible, we utilize the 2-minute cadence data processed by the Science Processing Operations Center (SPOC) pipeline, which provides high-quality background subtraction, systematic correction and flux-dilution corrections due to the large pixel scale of \tess\ \citep{kdph2020ksciApertures,kdph2020ksciCal,kdph2020ksciPA,kdph2020PDC}. A few of our candidate stars were only observed by \tess\ in the Full Frame Images (FFIs). For these targets, we only report on their variability if MAST hosts \tess-SPOC high-level science products \citep{Caldwell2020spoc}, since these are similarly processed by the SPOC pipeline and have corrections applied for crowding due to the large \tess\ pixels. These FFIs store data on the stars at a 30-minute or 10-minute cadence, depending on when the data were collected. See Table~\ref{tab:targets} for a list of those targets used in our analysis, including their TESS Input Catalog (TIC) identification and magnitude \citep{Stassun2018TIC}. All the data are available at MAST: DOI\, \dataset[10.17909/t9-tkr5-6a83]{\doi{10.17909/t9-tkr5-6a83}}.

Regardless of the data's cadence, we use the pre-search data conditioned, simple aperture photometry (PDCSAP) photometric time series \citep{kdph2020PDC} for our analysis. As an overview, the TESS pipeline that creates these light curves removes the background, performs simple aperture photometry using a fixed aperture that optimizes the signal-to-noise of the light curve, and then removes systematics by fitting Principal Component Analysis (PCA) basis vectors that represent common signals observed in nearby stars \citep{Smith2012,Stumpe2012PASP}. 

We then use the software \lightkurve\ \citep{lightkurve} to detrend and normalize the light curve. For some light curves we remove obvious systematic variations by applying a Savitzky-Golay filter \citep{savgol1964AnaCh..36.1627S} with a window length of half of the number of data points in the sector. As a sector is $\approx$27 days long, we only expect to be able to unequivocally measure variability with time scales shorter than half the length of the sector.  Variability longer than this time scale, unless very large in amplitude, is often mistaken for systematics and is commonly removed from the time series. While \tess\ data can be used to find longer-period trends, it requires very careful examination of possible systematic noise and stitching together consecutive sectors of data. For this study we restrict our analysis to repeatable variability of time scales shorter than $\approx$13\,days in length and large-amplitude, single-occurrence variations (such as eclipses or flares).

To look for periodic variations, we used Lomb-Scargle periodograms \citep{Lomb1976,lightkurve} and report those with significant peaks that otherwise do not appear to be caused by systematic noise (see Section~\ref{sec:var}).  For those with no significant variability, we provide upper limits to any variability that could be detected (see Section~\ref{sec:novar}).

\subsection{Stars with Variability} 
\label{sec:var}

\begin{figure*}
    \centering
    \includegraphics[width=0.9\linewidth]{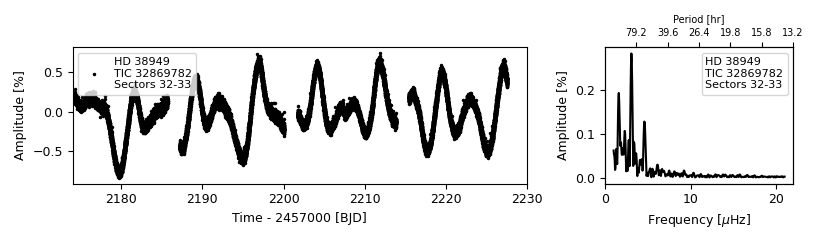}
    \includegraphics[width=0.9\linewidth]{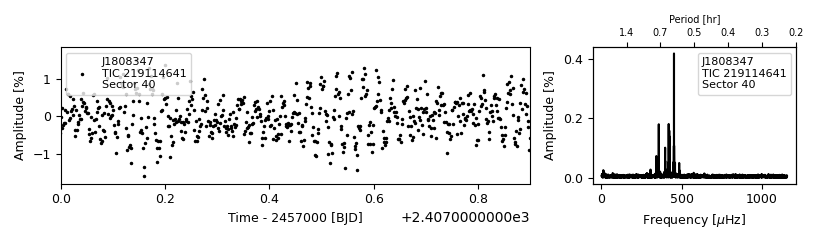}
    \includegraphics[width=0.9\linewidth]{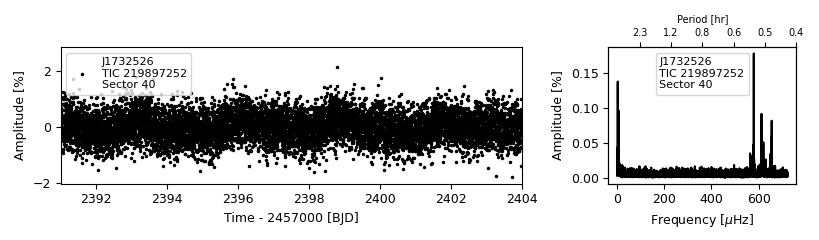}
    \includegraphics[width=0.9\linewidth]{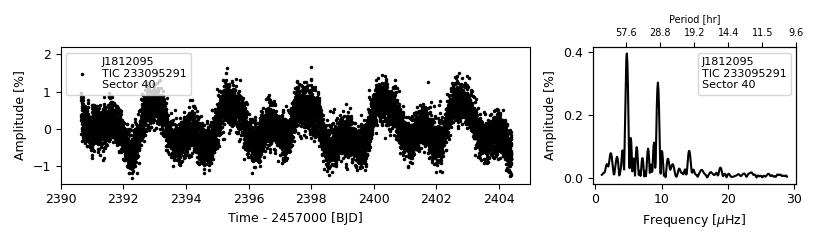}
    \caption{Light curve (left) and periodogram (right) of stars with peak to peak detected variability larger than approximately half a percent. The \tess\ sector(s) used to generate both are labeled on each plot.}
    \label{fig:lcft1}
\end{figure*}

\begin{figure*}
    \centering

    \includegraphics[width=0.9\linewidth]{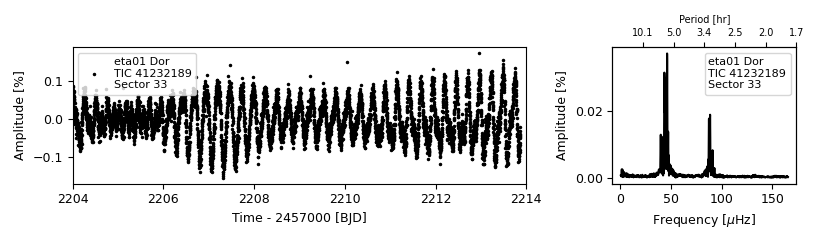}
    \includegraphics[width=0.9\linewidth]{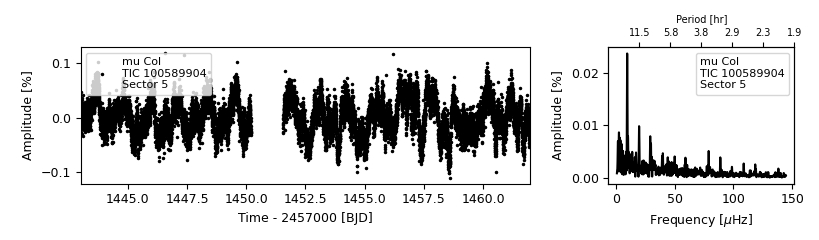}
    \includegraphics[width=0.9\linewidth]{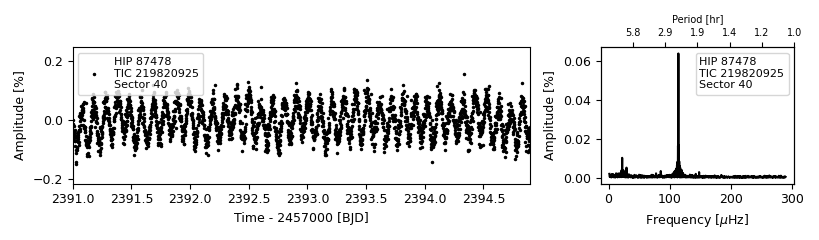}
    \includegraphics[width=0.9\linewidth]{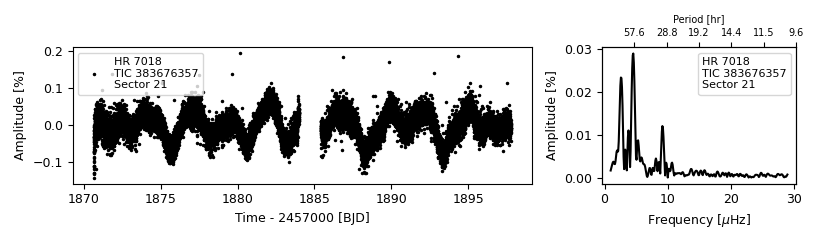}

    \caption{Light curve (left) and periodogram (right) of stars with significant detected variability. The \tess\ sector used to generate both are labeled on each plot.}
    \label{fig:lcft3}
\end{figure*}

\begin{figure*}
\centering
    \includegraphics[width=0.9\linewidth]{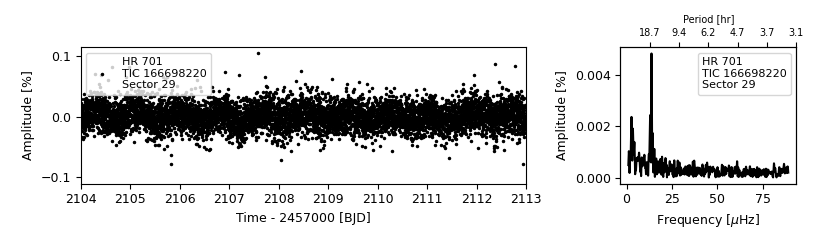}
    \includegraphics[width=0.9\linewidth]{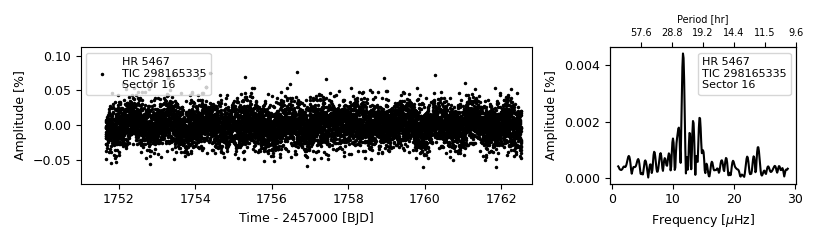}
    \includegraphics[width=0.9\linewidth]{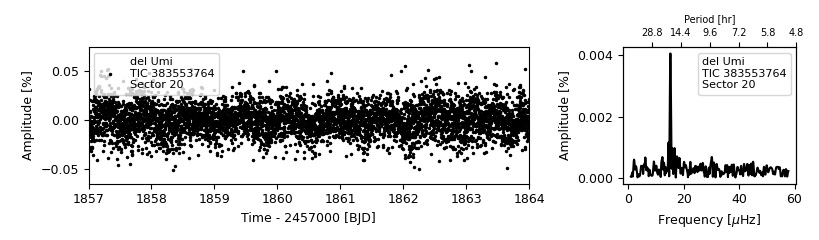}
    
    \caption{Light curve (left) and periodogram (right) of stars with significant detected variability. The \tess\ sector(s) used to generate both are labeled on each plot.}
    \label{fig:lcft4}
\end{figure*}

We found 15 of the \webb\ spectrophotometric standards stars in our sample to have statistically significant periodic variability. Only four show variability in flux larger than 1\% when considering the full extent of the observed variability (minimum to maximum brightness, or peak-to-peak).  For each variable star, the largest peak in the periodogram is highly significant and is at least 5 times larger than the noise level of the periodogram \citep{KjeldsenBedding1995, Baran2021}. Figures~\ref{fig:lcft1} -- \ref{fig:lcft4} show parts of the light curve and Lomb-Scargle periodogram created from one entire \tess\ sector for the 15 variable stars in our sample (two sectors for the case of HD~38949). 
Figure~\ref{fig:lcft1} highlights the four with the largest variations.

Table~\ref{tab:variable} summarizes the amplitude of the variations and their approximate period.  It lists the period and amplitude of the maximum amplitude peak in the periodogram. Since most stars have more than one period of variability, the maximum amplitude in the periodogram does not capture the full change in relative flux observed in the \tess\ light curves. To approximate full observed peak-to-peak variability in the presence of noise, we calculate the quantity $V_{95}$, which is defined to be the difference between the maximum and minimum values of those points contained by 95.45\% of the values in the light curve centered on the median.  $V_{95}$ measures the envelope of all data within $\pm 2\sigma$ of the median.

When reporting the statistics we use the sectors shown in Figures~\ref{fig:lcft1} -- \ref{fig:lcft4} and bin the light curves to 6-minute bins to improve the signal-to-noise ratio in some of the dimmer stars. The displayed sectors were chosen to represent the observed variability and have the lowest observed noise. The $V_{95}$ statistic acts as a reasonable measure of the full amplitude of the observed variability for those whose amplitudes are significantly larger than the noise.  For the fainter stars with low amplitudes and noisy light curves, $V_{95}$ overestimates the true extent of the observed variability. For example, for the dim variable star, J1732526 ($T_{mag}=12.5$), the precision on the flux (estimated from the average power in the periodogram, \citealt{KjeldsenBedding1995}) is 0.5\%, one fifth of the $V_{95}$ statistic for this star.  No single statistic will fully summarize the variability of these stars; the light curves and periodograms provide a more complete picture of the amplitude, timescale and shape of the observed periodicity.

A single \tess\ pixel covers more than 21\,arcsec of the sky on a side and the point spread function's full-width-half-max ranges from 1.13--2.76 pixels \citep{oelkers2018}, making it possible for nearby sources to contaminate our light curves \citep[such as that seen in exoplanet catalogs, see][]{Coughlin2014AJ}. As a way to quickly assess the impact of nearby sources, Table~\ref{tab:variable} includes the crowding metric calculated by the SPOC pipeline.  This metric indicates the fraction of the light in the aperture that comes from the target star given the positions and amplitudes of stars in the TIC.  A value near 1.0 indicates an isolated star, while lower values indicate significant crowding from neighbors.  Two of our large-amplitude variables (J1732526 and J1812095) have crowding values around 0.8.  The SPOC Pipeline corrects the light curve amplitudes for any dilution due to nearby stars by using this crowding value \citep{FausnaughDRN}.  As a result, the amplitudes and upper limits we report in this paper have already accounted for these neighbors.

\begin{figure*}
\centering
\includegraphics[width=0.45\linewidth]{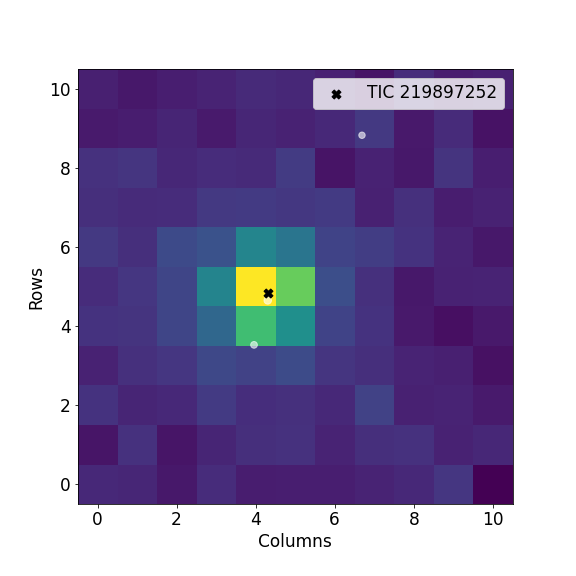}
\includegraphics[width=0.45\linewidth]{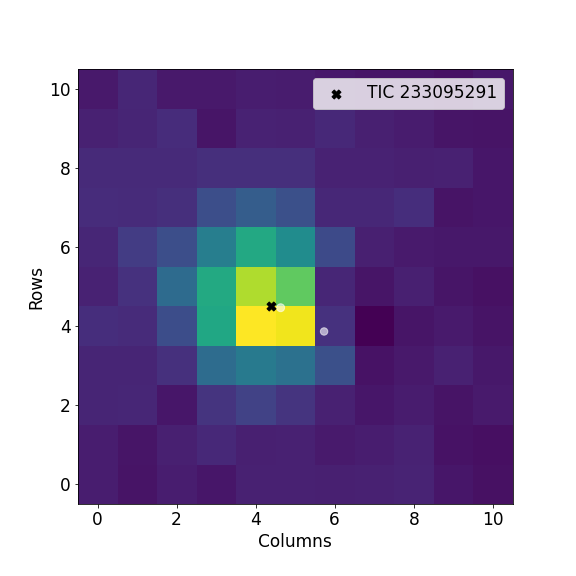}
\caption{Heat maps of the fitted amplitudes for each 21 arc\,second pixels at the observed periods across the pixels downloaded from \tess\ in Sector 40 for this target (generated by \tesslocalize\ \citealt{HiggensBell}.) TIC 219897252 (J1732526) is on the left and TIC~233095291 (J1812095) is on the right. The grey circles represent known Gaia Data Release 2 sources with $T_{mag} > 15$, approximately three magnitudes dimmer than the target stars. The black 'x' represents the best fit between the heat map and the \tess\ PRF. In both cases the best fit location overlaps the Gaia location of our target at the center of the \tess\ pixels, indicating that the variability comes from the targeted source.} 
\label{fig:cent}
\end{figure*}

Due to the presence of nearby stars for two of our large variables, we investigated the false positive scenario that one of the nearby stars is varying instead of the target of interest. To test this possibility, we use the code \tesslocalize\ \citep{HiggensBell} to determine if the location of the observed variations in the pixel time series is consistent with the position of the target star, as reported by Gaia Data Release 2 \citep{gaiaDR22018}.  We analyzed the large amplitude frequencies from Sector~40 data for both J1732526 and J1812095. The heat maps of per-pixel amplitudes shown in Figure~\ref{fig:cent} for J1732526 were calculated using the five largest frequencies and the two largest frequencies for J1812095 apparent in the periodogram.  These per-pixel amplitudes are then fit to the \tess\ pixel response function (PRF) to obtain the true location of the variability. In both cases the location on the CCD of the largest amplitudes overlaps with our target star. 

For completeness, we analyzed all 15 variable stars with the same technique, despite having no bright, nearby sources. In no case did evidence emerge that a nearby source caused the variations, though for stars with less significant variability, it was not always possible to convincingly extract and fit per-pixel variability.

\input{variable_stats}
\vspace{-1em}

\subsection{Reasons for Variability} 

While we do not attempt to definitively explain the physical reasons for the variability we observe in these stars, it is likely the reason for the photometric variability is related to either stellar pulsations or spots. 

Many of our variable stars are A dwarfs.  Variability in A-type stars can occur due to a plethora of reasons, including stellar pulsations.  The $\delta$ Scuti variables appear at the intersection of the instability strip and the main sequence \citep[e.g.,][]{Petit1987}.  They are the most prominent pulsators among A dwarfs, and they make up $\approx$27\% of all variable A and F dwarfs \citep{Uytterhoeven2011}. They pulsate in pressure modes and mixed pressure and low-order gravity modes due to the $\kappa$ mechanism \citep{Lee1985}, with pulsation frequencies in the range 18\,min to 8\,hrs \citep{Amado2004}. $\gamma$ Doradus variables also appear in a similar part of the HR diagram \citep[e.g.,][]{Kaye1999}, although they are rarer than $\delta$ Scuti stars, making up $\approx$12\% of all variable A and F dwarfs \citep{Uytterhoeven2011}. $\gamma$ Dor variables pulsate in non-radial, high-order gravity modes caused by convective blocking \citep{Guzik2000} with pulsation periods typically on the order of one day \citep{Grigahcene2010}.  Some stars, known as hybrid $\gamma$ Dor-$\delta$ Scuti stars, exhibit both $\gamma$ Dor and $\delta$ Scuti pulsations \citep{Grigahcene2010}.  These hybrid pulsators account for an additional $\approx$23\% of all variability found in A and F dwarfs. In our sample we find evidence of all three types of the above pulsators at varying amplitudes.

Other than pulsations and binarity can also cause variations in A stars. \citet{Duchene2013} determined the multiplicity fraction of stars with masses between 1.5 and 5.0\,{{$\rm M_{\odot}$}} to be $\ge$\,50\%. While short-period eclipsing binaries tend to be obvious in light curve data due to their significant flux variations during eclipse events, long-period eclipsing binaries and ellipsoidal variables are more difficult to identify. A comparison of the orbital period distribution identified in the Kepler Eclipsing Binary Catalog \citep{Kirk2016} with that of the \tess\ Eclipsing Binary Catalog \citep{Prsa2021} shows that the \tess\ Eclipsing Binary Catalog suffers from duty cycle suppression, which contributes to a decline in the identification of binary stars with orbital periods greater than 13 days \citep{Prsa2021}. While this comparison includes ellipsoidal variables (including eccentric ellipsoidal variables known as heartbeat stars; \citealt{Thompson2012}), their amplitudes are a function of their inclination and periastron distance. Thus, depending on these factors, ellipsoidal variations can be hidden in the noise, even at shorter periods. As our sample has been vetted using longer-baseline observations, as expected, we do not find any eclipses in the data. The increased precision of the \tess\ observations would allow the identification of lower-amplitude ($\sim$ few hundred parts per million) ellipsoidal variables if present, however, we do not find any evidence of ellipsoidal variations in our sample.


 Rotational modulation due to spots can also cause significant variations in the light curve. \citet{Balona2013, Balona2017} claim to have found spots in more than 50\% of A stars in the Kepler data, suggesting that modulation due to surface rotation is more commonly observed than previously thought. However, the origin of spots in A and B stars is currently under debate, as these stars do not have a significant surface convective zone, which is usually associated with surface spots. The presence of a fossil magnetic field \citep{Parker1955} and the creation of a magnetic field caused by a dynamo effect in mass motions that occur in convective layers \citep{Charbonneau2014} are two of the contending ideas for the creation of spots in hotter main sequence stars \citep{Balona2021}. In our sample, we find signatures that suggest the presence of spots in several objects:  HD~38949, J1812095, $\lambda$~Lep, $\eta^1$~Dor, and HR~7018.

\subsection{Large-Amplitude Variable Stars} 

In this section we discuss in more detail the amplitude and astrophysics for the variability of the six stars with the largest changes in brightness that we measured in the \tess\ data from our target list. Each of these stars shows a $V_{95}\ge0.35$\%. In these cases, the time scale and amplitude of the variations could impact high-precision spectrophotometric calibrations, and their variability should be considered before using them as a standard star. The remaining nine stars have small enough amplitudes that at least for JWST they will not significantly impact the calibrations. 

Three of the large-amplitude variable stars are A dwarfs from the \citet{Reach2005} Spitzer calibration. The amplitudes we see here were likely below the level of detection for Spitzer. For comparison, \citet{Krick2021IRAC} measured the IRAC fluxes in dozens to thousands of observations for most of our variable stars and reported standard deviations between 0.7\%\ and 2.8\%\ for the 3.6~$\mu m$ band.  This observed scatter is approximately four times larger than the variability we report here.

\subsubsection{HD~38949 (TIC~32869782)}

This G dwarf ($T_{mag}=7.3$) varies with a peak-to-peak amplitude of 1.2\% (Figure~\ref{fig:lcft1}, top panel). \tess\ observed HD~38949 in Sectors 6, 32, and 33.  The power spectrum is complex and shows a 7.9-day peak along with peaks at both a half and quarter of that period. This quasi-periodic, oscillatory behavior is consistent with a combination of stellar rotation at a 7.9-day period and migrating star spots on the surface due to strong magnetic activity \citep{Santos2021}.  HD~38949 is also an x-ray source, as revealed by the Swift Observatory \citep{Swift2020,Evans2020vizier}, which supports a picture of active spots and flares. Due to these properties, the \jwst\ calibration team has already removed this star from the list of standards \citep{Gordon2022inprep}.

It is also noteworthy that HD~38949 is listed in the Hubble Space Telescope Calspec database\footnote{\url{https://www.stsci.edu/hst/instrumentation/reference-data-for-calibration-and-tools/astronomical-catalogs/calspec}} \citep{Bohlin2014PASP126} as a star with complete STIS coverage that could be used to support flux calibrations of other observatories.

\vspace{2em}
\subsubsection{J1808347 (TIC~219114641)}

This A dwarf ($T_{mag}=11.9$) shows a rich set of peaks in its periodogram that produce a peak-to-peak amplitude of 1.65\% (Figure~\ref{fig:lcft1}, second panel). These variations compromise its ability to act as a spectrophotometric standard. The periods range from 0.5 to 1.0 hours, with the strongest mode at 0.6 hours.  The high-frequency variations are caused by $\delta$~Scuti pulsations.  The variability is difficult to see in the 30-minute \tess\ data taken during the first year of TESS observations; however, they are more apparent in the 2-minute cadence data from Sectors 40 and 41.  This star will be observed again by \tess\ in Sectors 47--55.

\subsubsection{J1732526 (TIC~219897252)}

J1732526 is an A dwarf ($T_{mag}=12.5$) that behaves much like J1808347; it is multiperiodic, with its strongest mode having a period of 0.5 hours, suggesting most of the variability arises from $\delta$ Scuti pulsations (Figure~\ref{fig:lcft1}, third panel). Its light curve has a peak-to-peak amplitude of roughly 1.4\%, limiting its value as a spectrophotometric standard. However, in this case, this statistic is somewhat inflated by the inherent noise of this dim object. The combined amplitudes of the largest peaks point to a peak-to-peak variation closer to 1\%.  In addition to the roughly 0.5-hour pulsation modes, the light curve also shows a longer-period component (roughly 2.5 days). This latter component could arise from rotation, or it could indicate that J1732526 is a hybrid $\delta$ Scuti ($p$-mode) -- $\gamma$ Doradus ($g$-mode) pulsator. \tess\ observed this star at a 2-minute cadence in Sectors 40 and 41 and is planning to observe it again in Sectors 48--55.

\subsubsection{J1812095 (TIC~233095291)}

This A dwarf ($T_{mag}=11.6$) shows consistent oscillations with periods of 1--3\,days and peak-to-peak changes in the flux greater than 1.5\% in the TESS bandpass (Figure~\ref{fig:lcft1}, bottom panel).  The light curve and periodogram show significant (though somewhat variable in amplitude) variations at a period of 2.44 days, with a secondary peak at exactly half that period, which indicates rotational modulation due to spots.  The periodogram shows no evidence for short-period peaks (i.e., for periods of an hour or less).  \tess\ observed this star in Sectors 14--26 at a 30-minute cadence, and at a 2-minute cadence in Sector 40.  The variability is approximately the same across all sectors. The large peak-to-peak variability on long time scales may make this target unsuitable for high-precision spectrophotometric calibration.

\subsubsection{HR~6514 (TIC~198456033)} 

The periodogram of this bright A dwarf ($T_{mag}=6.4$) shows a complex set of peaks with periods between roughly 0.5 and 2.0 hours, with up to 0.42\% peak-to-peak variability (Figure~\ref{fig:lcft2}, top panel).  The strongest mode has a period of 1.3 hours.  The periodicity is too fast to arise from stellar rotation and thus likely arises from short-period, $\delta$~Scuti ($p$-mode) pulsations.  \tess\ observed this star in Sectors 14--26 and 40--41, and is expected to observe it again in Sectors 47--55.  In the present data, the variability is consistent in amplitude from one sector to the next.

\subsubsection{$\lambda$ Lep (TIC~442871031)}

\begin{figure*}
\centering
    \includegraphics[width=0.9\linewidth]{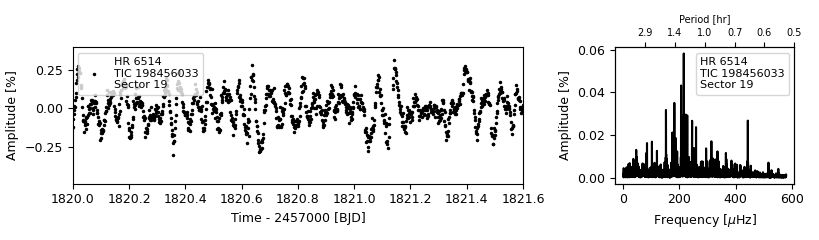}
    \includegraphics[width=0.9\linewidth]{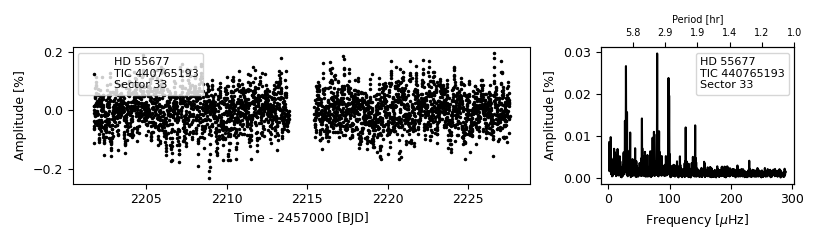}
    \includegraphics[width=0.9\linewidth]{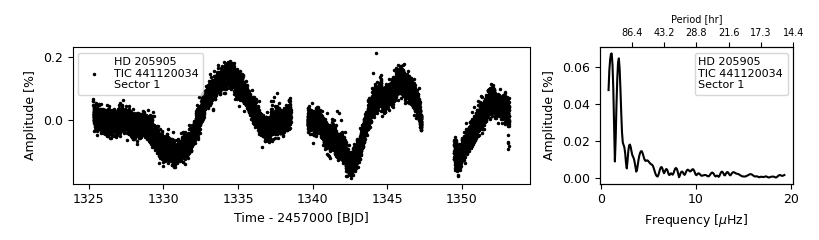}
    \includegraphics[width=0.9\linewidth]{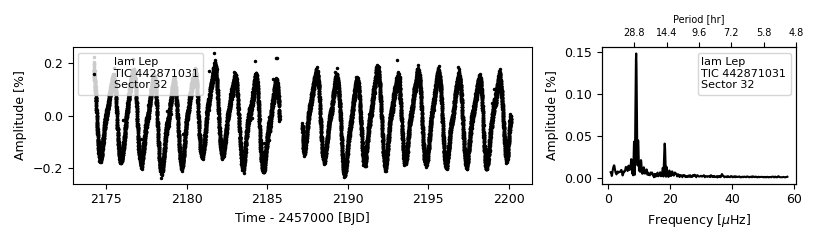}

    \caption{Light curve (left) and periodogram (right) of stars with significant detected variability. The \tess\ sector used to generate both are labeled on each plot.}
    \label{fig:lcft2}
\end{figure*}

This B9.5 dwarf ($T_{mag}=4.6$) varies with a coherent period of 1.26\,days and multiple, exact harmonics of this base period, which suggests that the variation is due to spots on the stellar surface. The peak-to-peak amplitude is greater than 0.3\% (Figure~\ref{fig:lcft2}, bottom panel). \tess\ observed this star in two sectors: 5 and 32. In both sectors the variability was similar in amplitude and shape. $\lambda$~Lep is one of the hot B dwarfs under consideration for use as standards starting in \jwst\ Cycle 2 \citep{Gordon2022inprep}.

\subsection{Not Observed to Vary} 
\label{sec:novar}

\begin{figure*}
    \centering
    \includegraphics[width=0.8\linewidth]{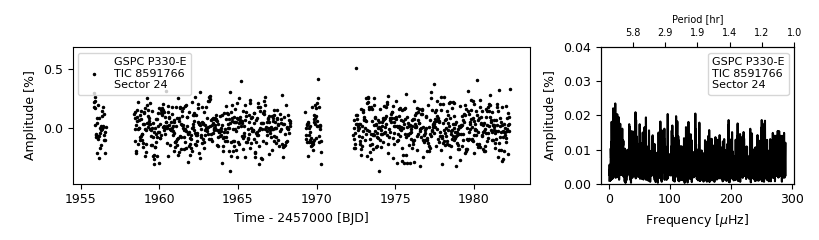}
    \includegraphics[width=0.8\linewidth]{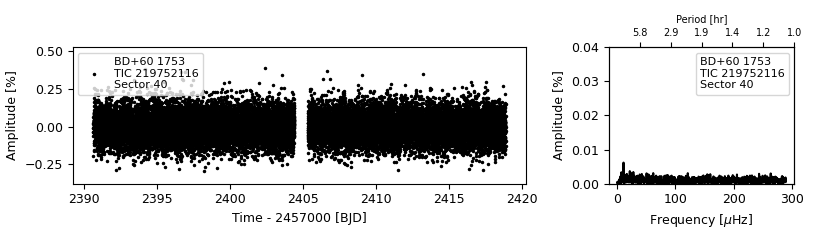}
    \includegraphics[width=0.8\linewidth]{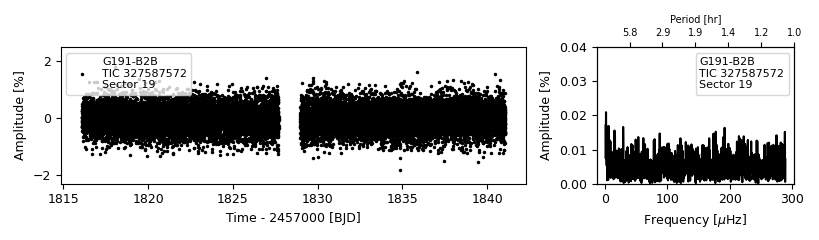}
    \caption{Light curves and Lomb-Scargle periodograms of three common calibration stars that show no significant, coherent variability in the \tess\ light curves.}
    \label{fig:novar}
\end{figure*}

We find no significant variability in 22 of the 37 stars we examined with existing \tess\ data. Figure~\ref{fig:novar} displays the light curves and periodograms for three well-known calibration stars with no significantly detected variability. We note that \tess\ commonly has large-amplitude systematic noise in its light curves at long periods, including noise at periods of 2~weeks due to the orbit of the spacecraft and near 1~day due to scattered light from the rotation of the Earth \citep{Luger2019}. Our sample of variable stars does not include those stars that only showed evidence of these types of systematic noise.

For those cases with no significant variability, we provide two statistics to better understand the upper limits of variability that could exist undetected in the \tess\ data in Table~\ref{tab:novar}. The photometric precision achieved by the \tess\ data is mostly driven by the brightness of the star, though in some cases systematic noise is playing a part. The quantity $A^{max}_{<1d}$ is the maximum amplitude peak in the periodogram seen at periods less than 1 day. This statistic can serve as measure of the upper limit on consistent, short-period variability. $V_{99.7\%}$ can be interpreted as the largest peak-to-peak variation on time scales longer than 30-minutes that could be present in the data without detection. It is calculated by binning the light curve to a 30-minute cadence and reporting the difference between the minimum and maximum relative flux for those points that lie within 99.7\% of the median observed flux. If the data are predominantly Gaussian noise, this value is approximately six times the size of the standard deviation of the light curve. For both of these statistics we report the sector that gives the smallest value, as some \tess\ sectors are noisier than others depending on scattered light from the Earth and Moon. 

\tess\ is able to provide limits to changes in brightness below 1\% for all but six stars, all of which are dimmer than $12.9$\,mag in the \tess\ band.  For all but our dimmest star, \tess\ sees no evidence of coherent, periodic variability with amplitudes larger than 0.2\% for periods shorter than 1 day. Thus, \tess\ can still confirm the suitability as flux standards of many of the faintest sources in our sample.

\input{novar_stats}
\vspace{-1em}

\subsection{Limits on using TESS data to finding variability}

Because variability at long periods due to scattered light from the Earth-Moon system can produce false positives \citep{Luger2019}, our TESS observations are only sensitive to relatively short-term variability with periods less than two weeks. We are confident that our detections of variability are not instrumental in nature given the short periods observed, and the fact that many are multi-periodic pulsators with many oscillation modes detected with high significance. 

For the variable stars found in our sample, the possibility that the detected variability comes from another source is small, even for the low-amplitude stars where the localization of the signal could not be well constrained (see Section \ref{sec:var}). The variable stars reported here are uncrowded. All have TIC CROWDSAP values exceeding 0.75 (with most exceeding 0.99), indicating that the targets themselves contribute more than 75\% of the flux in the extracted TESS aperture. If variability were found in dimmer stars in more crowded fields, this type of false positive would be a larger concern.

While targets with multi-sector TESS data could have revealed some types of long-period variability, such as eclipses from binary systems, the data do not rule-out most variability on time scales of several weeks or longer. For example, spotted stars \citep{Mcquillan2014} and eclipsing (or ellipsoidal) binaries \citep{Prsa2021} have been found with periods longer than a few weeks. Other facilities would be required to empirically test for variability on longer timescales. For example, extensive ground-based campaigns have been undertaken to monitor the absolute fluxes of hundreds of Gaia spectrophotometric standard stars (e.g., \citealt{2016MNRAS.462.3616M,2021MNRAS.501.2848A}). The empirical photometric uncertainties of Gaia itself could also reveal long-term variable stars unsuited to be flux standards (e.g., \citealt{2021ApJ...912..125G,2021ApJ...908..180A}).

As there are no long-period pulsators amongst A stars, we anticipate that our detection of pulsations for that group is complete within our sample of stars to a precision of a few hundred parts per million. The one exception is the four objects extracted from the FFIs, which could have undetected high-frequency pulsations.

\section{Conclusions} 
\label{sec:conclusion}

The high-cadence, all-sky photometric survey of bright stars produced by \tess\ is a useful resource to vet potential spectrophotometric calibration stars for variability on time scales of a few minutes to a couple of weeks. The precise relative photometry, month-long stares, and fast cadence provided by \tess\ is a significant improvement over previous efforts to vet these stars and has revealed peak-to-peak changes in flux as large as 1.4\% for four of the \webb\ candidate calibrators. Another 11 show lower-amplitude variability. This variability appears to arise from either stellar pulsations or stellar spots rotating in and out of view. At the same time, relative photometry from \tess\ has set upper limits on optical to near-infrared variability for most of the candidate standard stars at a level well below that needed to achieve the spectrophotometric requirements for \jwst.  Identifying those with known issues provides the \webb\ calibration team with information necessary to pick those stars with the highest likelihood of accurate calibrations in the least amount of observing time. This work has already led to revisions of the \webb\ spectrophotometric calibration star list; see \citet{Gordon2022inprep}.

Some of the stars vetted in this paper are the same stars that were used to calibrate telescopes such as Spitzer \citep{Reach2005} and Hubble \citep{Bohlin2011AJ}, and some may be used to calibrate future space and ground-based telescopes.  Because \tess\ is an ongoing all-sky survey, in the future \tess\ will serendipitously continue to monitor many of the \webb\ spectrophotometric calibration stars. This monitoring by \tess\ will reveal if the star varies photometrically in new or unexpected ways.  Some of these observations may even be contemporaneous with the \webb\ calibration observations, as both missions plan to be operating at the same time.  This work provides further evidence that small NASA missions like \tess\ can provide crucial supporting observations which enable larger missions like \jwst\ to accomplish their science goals.

\begin{acknowledgments}
This paper includes data collected by the \tess\ mission. Funding for the \tess\ mission is provided by NASA's Science Mission Directorate. We thank the STScI internship program and \jwst\ for providing funding for MK. JJH acknowledges financial support through NASA Award 80NSSC20K0592. This research has made use of the VizieR catalogue access tool, CDS, Strasbourg, France (DOI: 10.26093/cds/vizier). The original description of the VizieR service was published in A\&AS 143, 23. This research has been made possible by the MAST archive for access to the \tess\ light curves, target pixel files, and \tess\ Input Catalog. The authors thank the referee for providing valuable comments that improved the scope of the paper.
\end{acknowledgments}

\vspace{5mm}
\facilities{TESS, JWST, MAST}

\software{astropy \citep{astropy2013,astropy2018},  
          lightkurve \citep{lightkurve},
          TESS\_Localize \citep{HiggensBell}
          }

\bibliography{references}{}
\bibliographystyle{aasjournal}

\end{document}

%% file: targetTable.tex
\startlongtable
\begin{deluxetable*}{lrcrcr} 
\tablecolumns{6}
\tablewidth{0pt}
\tablecaption{Candidate standard star properties with TESS data. }
\tablehead{
  \colhead{Target Name } & \colhead{TIC ID}  & \colhead{Class} & \colhead{T}& 
  \colhead{Cadence}& \colhead{Coord.}\\
  \colhead{} & \colhead{} & \colhead{}& \colhead{[mag]}& \colhead{min} & \colhead{J2000}
}
\startdata
GSPC P330-E &   8591766 & (G0 V) & 12.4 &  30  & 16 31 34  +30 08 46  \\
   16 Cyg B &  27533327 &         G3 V &  5.6 &   2  & 19 41 52  +50 31 03  \\
   HD 38949$^a$ &  32869782 &         G1 V &  7.3 &   2  & 05 48 20  $-$24 27 50  \\
    HD 6538 &  39464221 &         G1 V &  5.9 &   2   & 17 32 01  +34 16 16  \\
  $\eta^1$\,Dor &  41232189 &         A0 V &  5.8 &   2   & 06 06 09  $-$66 02 23  \\
      $\mu$ Col & 100589904 &       O9.5 V &  5.5 &   2  & 05 46 00  $-$32 18 23  \\
     10 Lac & 128692445 &         O9 V &  5.1 &   2   & 22 39 16  +39 03 01  \\
   HD 37962 & 140282069 &         G2 V &  7.2 &   2    & 05 40 52  $-$31 21 04  \\
WD 1057+719 & 147921014 &        DA1.2 & 15.1 &   2    & 11 00 34  +71 38 03  \\
     GD 153 & 149505899 &        DA1.2 & 13.7 &   2    & 12 57 02  +22 01 53  \\
  HD 116405 & 165370459 &         A0 V &  8.4 &   2    & 13 22 45  +44 42 54  \\
     HR 701 & 166698220 &         A5 V &  5.7 &   2    & 02 22 55  $-$51 05 32  \\
  HD 101452 & 181240911 & (A9 V) &  7.3 &   2    & 11 40 14  $-$39 08 48  \\
    HR 6514 & 198456033 &         A4 V &  6.4 &   2    & 17 26 05  +58 39 07  \\
   J1757132$^*$ & 219094190 & (A9 IV) & 11.6 &  30    & 17 57 13  +67 03 41  \\
GSPC P041-C$^b$ & 219015049 &       (G0 V) & 11.5 &   2    & 14 51 58  +71 43 17  \\
   J1808347$^*$ & 219114641 &       (A3 V) & 11.9 &   2    & 18 08 35  +69 27 29  \\
 BD+60 1753 & 219752116 &         A1 V &  9.7 &   2    & 17 24 52  +60 25 51  \\
  HD 163466 & 219820925 & (A7 V) &  6.7 &   2    & 17 52 25  +60 23 47  \\
   J1732526$^*$ & 219897252 &       (A4 V) & 12.5 &   2    & 17 32 53  +71 04 43  \\
  HD 180609 & 229945862 & (A3 V) &  9.3 &   2    & 19 12 47  +64 10 37  \\
  HD 115169$^c$ & 229980646 &         G3 V &  8.7 &   2    & 13 15 47  $-$29 30 21  \\
   J1802271$^*$ & 233067231 &       (A2 V) & 12.0 &   2    & 18 02 27  +60 43 36  \\
   J1805292$^*$ & 233075513 & (A3 V) & 12.2 &   2    & 18 05 29  +64 27 52  \\
   J1812095$^*$ & 233095291 &       (A3 V) & 11.6 &   2    & 18 12 10  +63 29 42  \\
   J1743045$^*$ & 233205654 & (A8 V) & 13.3 &   2    & 17 43 04  +66 55 02  \\
      GD 71 & 247923021 & DA1.5 & 13.4 &   2    & 05 52 28  +15 53 13  \\
    HR 5467 & 298165335 &         A1 V &  5.8 &   2    & 14 38 15  +54 01 24  \\
   G191-B2B & 327587572 &        DA0.8 & 12.2 &   2    & 05 05 31  +52 49 52  \\
  HD 167060 & 365653206 &         G3 V &  8.4 &   2    & 18 17 44  $-$61 42 32  \\
     $\delta$ UMi & 383553764 &       A1 Van &  4.4 &   2    & 17 32 13  +86 35 11  \\
    HR 7018$^d$ & 383676357 &         A0 V &  5.8 &   2    & 18 37 34  +62 31 36  \\
GSPC P177-D & 417544924 & (G0 V) & 12.9 &  30    & 15 59 14  +47 36 42  \\
   HD 55677 & 440765193 &         A2 V &  9.4 &  10    & 07 14 31  +13 51 37  \\
  HD 205905$^e$ & 441120034 &    G1.5 IV-V &  6.2 &   2    & 21 39 10  $-$27 18 24  \\
   $\lambda$ Lep & 442871031 &      B0.5 IV &  4.6 &   2    & 05 19 35  $-$13 10 36  \\
 WD1657+343 & 471015233 &        DA0.9 & 15.8 &   2    & 16 58 51  +34 18 53  \\
\enddata

\tablecomments{Spectral types in parenthesis are not optical spectral classifications and instead are based on photometry and/or spectra from STIS on Hubble.  References for the spectral types are the same as those specified by \citet{Gordon2022inprep}. References for the remaining spectral types is as follows: $^a$\citet{Houk1988mcts}, $^b$\citet{Bohlin2011AJ}, $^c$\citet{Houk1982mcts-hd115169}, $^d$\citet{Crowley1969AJ}, $^e$\citet{Keenan1989ApJS}. \newline
$^*$ Stars with names beginning with 'J' are Spitzer standards described in \citet{Reach2005} and as done their are given shortened names based on the 2MASS designation. J1757132 is 2MASS~J17571324+6703409. J1802271 is 2MASS J18022716+6043356. J1732526 is 2MASS~J17325264+7104431.  J1805292 is 2MASS~J18052927+6427520. J1808347 is 2MASS J18083474+6927286. J1812095 is 2MASS~J18120957+6329423. J1743045 is 2MASS~J17430448+6655015.}
\label{tab:targets}
\end{deluxetable*}

%% file: variable_stats.tex
\begin{deluxetable*}{lllcrcc}
\tablecolumns{6}
\tablewidth{0pt}
\tablecaption{Candidate spectrophotometric standards with observed variability.}

\tablehead{
  \colhead{Target Name} & \colhead{TIC ID} & \colhead{Class} & \colhead{CROWD$^{\dagger}$} & \colhead{Period} & \colhead{max Amp.} & \colhead{$V_{95}$} \\
  \colhead{} & \colhead{} & \colhead{} & \colhead{} &\colhead{[days]}& \colhead{[\%]}& \colhead{[\%]}
}
\startdata
 \object{HD 38949} &  32869782 &    G1V &    0.998 & 3.798 & 0.284 & 1.17 \\
\object[eta01 dor]{$\eta^1$\,Dor} &  41232189 &    A0V &    1.000 & 0.249 & 0.037 & 0.18 \\
   \object{$\mu$ Col} & 100589904 &    O9V &    0.999 & 1.196 & 0.024 & 0.12 \\
   \object{HR 701} & 166698220 &    A6V &    1.000 & 0.845 & 0.005 & 0.05 \\
 \object{HR 6514} & 198456033 &    A3V &    1.000 & 0.054 & 0.058 & 0.42 \\
 \object[2MASS J18083474+6927286]{J1808347} & 219114641 &    A3V &    0.991 & 0.026 & 0.419 & 1.65 \\
\object{HD 163466} & 219820925 &    A6V &    1.000 & 0.101 & 0.064 & 0.17 \\
 \object[2MASS J17571324+6703409]{J1732526} & 219897252 &    A3V &    0.831 & 0.020 & 0.178 & 1.40 \\
 \object[2MASS J18120957+6329423]{J1812095} & 233095291 &    A3V &    0.780 & 2.442 & 0.396 & 1.57 \\
  \object{HR 5467} & 298165335 &    A1V &    1.000 & 0.995 & 0.004 & 0.05 \\
 \object[del umi]{$\delta$ UMi} & 383553764 &    A1V &    1.000 & 0.761 & 0.004 & 0.04 \\
  \object{HR 7018} & 383676357 &    A0V &    0.996 & 2.547 & 0.029 & 0.13 \\
 \object{HD 55677} & 440765193 &    A4V &    0.984 & 0.145 & 0.029 & 0.24 \\
\object{HD 205905} & 441120034 &    G2V &    0.999 & 9.986 & 0.068 & 0.26 \\
 \object[lam Lep]{$\lambda$ Lep} & 442871031 &  B0.5V &    1.000 & 1.260 & 0.148 & 0.35 \\
\enddata

\tablecomments{The column Period gives the period of the largest amplitude peak (max Amp.) seen in the periodogram. $V_{95}$ is the peak-to peak range encompassing 95.45\% of the binned observed relative fluxes. Errors on the max Amp. value are less than 0.002\% with the exception being the 'J' stars ($T_{mag}\ge11$) whose errors are near 0.01\%.
\\
$^{\dagger}$CROWD comes from the CROWDSAP estimate from the \tess\ SPOC pipeline and is the fraction of flux from the target in the extracted aperture.}
\label{tab:variable}
\end{deluxetable*}

%% file: novar_stats.tex
\begin{deluxetable*}{llclcrr}
\tablecolumns{6}

\tablecaption{Upper limits on variability for those with no significant variations.}

\tablehead{
  \colhead{Name} & \colhead{TIC ID} & \colhead{Class} &\colhead{$T$} &\colhead{CROWD$^{\dagger}$} & \colhead{$A^{max}_{< 1d}$} & \colhead{$V_{99.7}$}\\
  \colhead{} & \colhead{} & \colhead{}& \colhead{[mag]} &\colhead{}& \colhead{[\%]}& \colhead{[\%]}
}
\startdata
GSPC P330-E &   8591766 &    G2V & 12.4 &    0.891 & 0.021 &  0.76 \\
   16 Cyg B$^*$ &  27533327 &    G3V &  5.6 &    0.630 & 0.017 &  0.21 \\
    HD 6538 &  39464221 &    G1V &  5.9 &    0.998 & 0.003 &  0.09 \\
     10 Lac & 128692445 &    O9V &  5.1 &    0.993 & 0.007 &  0.23 \\
   HD 37962 & 140282069 &    G2V &  7.2 &    1.000 & 0.002 &  0.09 \\
WD 1057+719 & 147921014 &  DA1.2 & 15.1 &    0.919 & 0.148 &  6.05 \\
      GD153 & 149505899 &    DA1 & 13.7 &    0.987 & 0.062 &  2.44 \\
  HD 116405 & 165370459 &    B9V &  8.4 &    0.969 & 0.003 &  0.10 \\
  HD 101452 & 181240911 &  A9mIV &  7.3 &    0.993 & 0.003 &  0.09 \\
   J1757132 & 219094190 &    A3V & 11.6 &    0.963 & 0.010 &  0.37 \\
 GSPC P041C & 219015049 &     G0 & 11.5 &    0.995 & 0.011 &  0.44 \\
 BD+60 1753 & 219752116 &    A1V &  9.7 &    0.992 & 0.006 &  0.17 \\
  HD 180609 & 229945862 &    A3V &  9.3 &    0.984 & 0.005 &  0.13 \\
  HD 115169 & 229980646 &    G2V &  8.7 &    0.984 & 0.003 &  0.11 \\
   J1802271 & 233067231 &    A3V & 12.0 &    0.963 & 0.023 &  0.82 \\
   J1805292 & 233075513 &    A1V & 12.2 &    0.965 & 0.014 &  0.49 \\
   J1743045 & 233205654 &    A5V & 13.3 &    0.986 & 0.032 &  1.32 \\
       GD71 & 247923021 &    DA1 & 13.4 &    0.788 & 0.055 &  1.90 \\
   G191-B2B & 327587572 &    DA0 & 12.2 &    0.903 & 0.021 &  0.74 \\
  HD 167060 & 365653206 &    G2V &  8.4 &    0.994 & 0.002 &  0.09 \\
GSPC P177-D & 417544924 &    G2V & 12.9 &    0.942 & 0.028 &  1.00 \\
 WD1657+343 & 471015233 &    DA1 & 15.8 &    0.631 & 1.289 & 37.88 \\
\enddata

\tablecomments{$^*$16 Cyg B is a known solar-type variable with low-amplitude modes (smaller than the limits reported here) at periods near 8 minutes (see \citealt{Metcalfe2012_16Cyg}). $A^{max}_{< 1d}$ is the maximum amplitude peak seen in the periodogram less than 1 day. $V_{99.7}$ is peak to peak flux change of those relative fluxes that lie within 99.7\% of the median flux.\\
$^{\dagger}$CROWD comes from the CROWDSAP estimate from the \tess\ SPOC pipeline and is the fraction of the flux that comes from the target star in the extracted aperture, as opposed to nearby sources in the TIC.}
\label{tab:novar}
\end{deluxetable*}